\documentclass[prl,twocolumn,aps,showpacs,superscriptaddress]{revtex4-1}
\usepackage{amsmath}
\usepackage{amssymb}
\usepackage{graphicx}
\usepackage{braket}
\usepackage[colorlinks=true,linkcolor=blue,citecolor=blue,urlcolor=blue]{
hyperref}
\usepackage{txfonts}

\newcommand{\E}{\mathrm{e}}
\newcommand{\im}{\mathrm{i}}
\newcommand{\vb}{\boldsymbol}

\newcommand{\idm}{\mathbb{I}}
\DeclareMathOperator{\tr}{tr}
\DeclareMathOperator{\dif}{d}
\DeclareMathOperator{\diag}{diag}

\begin{document}

\title{Robust asymptotic entanglement under multipartite collective dephasing}
\author{Edoardo G. Carnio}
\email{e.carnio@warwick.ac.uk}
\affiliation{Physikalisches Institut, Albert-Ludwigs-Universit\"at Freiburg,
Hermann-Herder-Stra\ss e 3, 79104 Freiburg, Germany}
\affiliation{Department of Physics, University of Warwick, Coventry, CV4 7AL,
United Kingdom}
\author{Andreas Buchleitner}
\affiliation{Physikalisches Institut, Albert-Ludwigs-Universit\"at Freiburg,
Hermann-Herder-Stra\ss e 3, 79104 Freiburg, Germany}
\affiliation{Freiburg Institute for Advanced Studies,
Albert-Ludwigs-Universit\"at Freiburg, Albertstra\ss e 19, 79104 Freiburg,
Germany}
\author{Manuel Gessner}
\email{manuel.gessner@physik.uni-freiburg.de}
\affiliation{Physikalisches Institut, Albert-Ludwigs-Universit\"at Freiburg,
Hermann-Herder-Stra\ss e 3, 79104 Freiburg, Germany}

\date{\today}


\begin{abstract}
We derive an analytic solution for the ensemble-averaged collective dephasing
dynamics of $N$ noninteracting atoms in a fluctuating homogeneous external field.
The obtained Kraus map is used to specify families of states whose entanglement properties 
are preserved at all times under arbitrary field orientations, even for states undergoing incoherent evolution. 
Our results apply to arbitrary spectral distributions of the field fluctuations.
\end{abstract}

\maketitle

Control of the coherent evolution of quantum systems in noisy environments
\cite{coherentev} is one of the crucial prerequisites for exploiting nontrivial
quantum effects in composite systems of increasing complexity. Whether in the
context of controlled molecular reactions \cite{molreac}, of many-particle
quantum dynamics \cite{platzer10}, or of quantum computers and simulators
\cite{nielsenchuang}, uncontrolled fluctuations and noise are detrimental to
most purposes of optimal control. Various strategies may be followed to
counteract the harmful influence of the environment: shielding the system
degrees of freedom \cite{haroche}, correcting environment-induced errors
\cite{QEC}, exploiting basins of attraction in dissipative systems \cite{Kraus}, or compensating dissipation---e.g., by coherent dynamics \cite{simeon}, dynamical decoupling \cite{DD}, or periodic measurements \cite{Zeno}. Such approaches can effectively reduce the environmental effects and can enhance coherence times, but a perfect protection of the quantities of interest is generally not possible.

By restricting to superposition states within a decoherence-free subspace, initially entangled states can be shielded completely from collective noise sources, hence protecting their entanglement at all times \cite{palma,DFS}. However, such subspaces are rather fragile to small perturbations, which limit their applicability in the context of dynamical processes \cite{Lidar}. Here, we identify conditions that ensure complete preservation of arbitrary degrees of entanglement, even for states that are not invariant under an incoherent time evolution. Specifically, we consider an important class of environment-induced fluctuations, which
are frequently encountered in state-of-the-art experiments \cite{Gross,HARTMUTREVIEW,Schindler}:
they manifest in intensity fluctuations of spatially homogeneous experimental
control fields, giving rise to an effective dephasing process. We show how control of the external field's orientation can lead to the complete
preservation of entanglement in bipartite---as well as multipartite---settings,
for arbitrary spectral characteristics of the control field fluctuations. We further identify families of states exhibiting time-invariant entanglement for arbitrary orientations of the external field.

To set the stage, let us consider a collection of $N$ noninteracting atomic
two-level systems with identical energy splitting $\hbar\omega$ controlled,
e.g., by a homogeneous magnetic field. Integration over the unavoidable
fluctuations of the latter's strength will induce a probability distribution
$p(\omega)$ of the characteristic energy splitting, and the $N$-atom quantum
state at time $t$ therefore needs to be described by the statistical operator 
\begin{align}\label{eq.separablemap}
\rho(t)=\int  p(\omega)U_\omega(t)^{\otimes
N}\rho(0)U^{\dagger}_\omega(t)^{\otimes N} \dif \omega,
\end{align}
provided the field fluctuations occur on time scales which are longer than the
time $t$ over which the $N$-atom state is propagated by the unitary
$U_\omega(t)^{\otimes N}$. In order to assess the open system time evolution of the quantum
correlations inscribed into the $N$-atom system, it is convenient to derive an
explicit expression for $\rho(t)$ in terms of the spectral distribution
$p(\omega)$ characterizing the fluctuations.

The single-atom propagator $U_\omega(t) = \E^{-\im H_{\omega} t/\hbar}$ is
generated by the time-independent single-atom Hamiltonian $H_{\omega} =(\hbar
\omega/2) \vb n \cdot \vb \sigma$, with $\vb
\sigma=(\sigma_x,\sigma_y,\sigma_z)$ being the vector of the Pauli matrices and $\vb n$
the orientation of the field. $H_{\omega}$ describes atomic dipoles interacting
with electromagnetic fields, as, e.g., the electronic qubits in trapped-ion
quantum registers \cite{HARTMUTREVIEW,Ben}. Introducing pairs of orthogonal
projectors $\Lambda_{\pm} = (\idm_2 \pm {\vb n \cdot \vb \sigma} )/2$, we can
rewrite the time evolution operator for a collection of $N$ atoms as
\begin{align}
U_\omega(t)^{\otimes N} & = \left( \E^{- \im \omega t/2} \Lambda_+ + \E^{\im
\omega t/2} \Lambda_- \right)^{\otimes N} \nonumber \\
 & = \sum_{j=0}^N \E^{\im \omega t \left(j - N/2 \right)} \Theta_j \, ,
\end{align}
where we have defined the operators
\begin{equation}
\Theta_j = \frac{1}{j! (N-j)!} \sum_{s \in \Sigma_N} V_s \left[
\Lambda_-^{\otimes j} \otimes \Lambda_+^{\otimes N-j} \right] V_s^{\dagger},
\end{equation}
where $\Sigma_N$ denotes the symmetric group and $V_s =\sum_{i_1\dots i_N}|i_{s(1)}\dots i_{s(N)}\rangle\langle i_1\dots
i_N|$ represents the permutation $s \in \Sigma_N$ in the operator space of $N$ qubits. The
ensemble-averaged state after time $t$,
\begin{equation}\label{eq.timeevolution}
\rho (t)  = \sum_{j,k=0}^N M_{jk}(t)   \Theta_j \rho(0) \Theta_k,
\end{equation}
is then fully characterized by the Toeplitz matrix $M(t)$, whose elements
$M_{jk}(t)=\varphi[(j-k)t]$ are generated by the characteristic function
$\varphi(t) = \int p(\omega) \E^{\im \omega t} \dif \omega$ of the probability
distribution $p(\omega)$. Bochner's theorem \cite{rudin} ensures that $M(t)$ is
a Hermitian semipositive definite matrix for all $t$. Diagonalization leads to
the canonical Kraus form \cite{Zyczkowski}
\begin{align}
\rho(t) = \epsilon_{t,0} \left[ \rho(0) \right] = \sum_{i=0}^N A_i(t) \rho(0)
A_i^\dagger (t),
\label{dephasing_map}
\end{align}
where the Kraus operators $A_i (t) = \sum_{j=0}^N  \sqrt{\lambda^i (t)} 
\lambda^i_j(t) \Theta_j$ contain the eigenvalues $\lambda^i(t)$ and the
components of the eigenvectors $\vb \lambda^i(t)$ of $M(t)$. Note, from the
structure of $A_i(t)$, that the above defined Kraus operators mediate an
effective interaction between the individual qubits---with its origin in the
spatial homogeneity of the external field. These environment-induced
interactions are able to create discord-type quantum correlations \cite{Ben}
and, as we will show in this Letter, given the appropriate control of $\vb n$, can
uphold multipartite entanglement at all times for arbitrary intensity
fluctuations.

Using the fact that the operators $\Lambda_\pm$ are orthogonal projectors in
$\mathbb{C}^2$, we can immediately show that both of the operators $\lbrace
\Theta_j \rbrace_j$ and $\lbrace A_i(t) \rbrace_i$ satisfy the condition $\sum_i
A_i^\dagger (t) A_i(t) = \sum_i \Theta_i^\dagger  \Theta_i = \idm_2^{N}$, which
ensures that the map $\epsilon_{t,0}$ defined in  \eqref{dephasing_map}---from
now on called the ``collective dephasing'' map---is not only completely
positive but also trace preserving for all $t$ \cite{Zyczkowski,Supp}. 

For absolutely continuous distribution functions \cite{abscont}, the
characteristic function vanishes asymptotically; i.e., $\lim_{t\rightarrow
\infty}\varphi(t)=0$. We  then have that $\lim_{t \rightarrow \infty} M(t) =
\idm_{N+1}$, and thus the Kraus operators reduce to $\lim_{t \rightarrow \infty}
A_i(t) = \Theta_i$. The asymptotic $N$-qubit state is thus given by $\rho_s =
\lim_{t \rightarrow \infty} \rho(t) = \sum_i \Theta_i \rho(0) \Theta_i $.
Because the operators $\Theta_i$ depend exclusively on the magnetic field
direction $\vb n$, the latter completely determines the properties of the
asymptotic state.

\begin{figure}[tb]
\centering
\includegraphics[width=.48\textwidth]{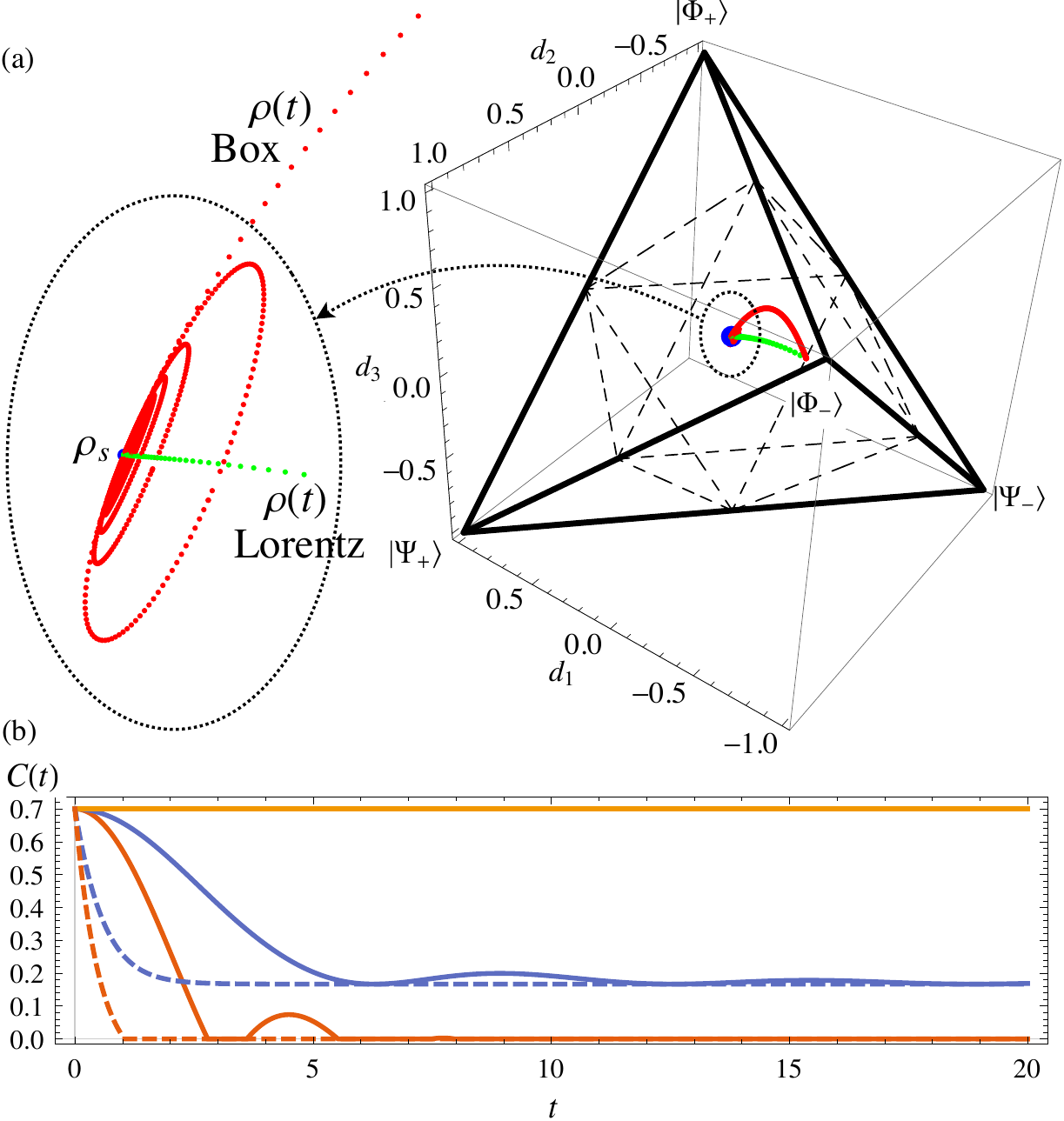}
\caption{Dynamics of a bipartite system with the initial state
$\rho_0 = \idm_4/20  + 4 \ket{\Phi^-}\bra{\Phi^-}/5 $. (a) The evolution of the
state under Lorentzian- (green) and box-distributed (red) noise,
$C_{0,1}(\omega)$ and $B_{0,1}(\omega)$ (see the
text for a definition), respectively, with a magnetic field direction $\vb n = (2,1,1)^T/
\sqrt{6}$, is depicted inside the tetrahedron of Bell-diagonal states. The inner
octahedron marks the set of separable states. (b) Depending on the orientation
of the magnetic field, the concurrence remains constant [for $\vb n=(1,0,0)^T$,
orange], decays to a finite value [for $\vb n = (2,1,1)^T / \sqrt{6}$, blue], or
decays to zero (for $\vb n = (0,1,1)^T / \sqrt{2}$, red). The decay is monotonic
for Lorentzian (dashed lines), or nonmonotonic for box-distributed (solid lines)
noise.}
\label{fig.dynamics}
\end{figure}

To gain some intuition on the time evolution of the entanglement properties,
e.g., of an $N$-ion quantum register under the action of the collective
dephasing map, we first consider two-qubit states with maximally mixed reduced
density matrices (also called Bell-diagonal states). Such states allow for a
simple geometric representation, since they are fully characterized by the
matrix $\beta_{ij} = \tr(\rho \cdot \sigma_i \otimes \sigma_j)$
\cite{Horodecki:1996gf}. There always exist unitary operations $U_A$ and $U_B$
such that $U_A \rho U_B^\dagger$ has a diagonal $\beta$ matrix, $\beta = \diag
\left(d_1, d_2, d_3 \right)$, while $\rho$ and $U_A\rho U_B^\dagger$ have the
same separability properties \cite{footnote1}. This allows us to associate with each density matrix a
point $\vb d = (d_1,d_2,d_3)^T \in \mathbb{R}^3$ and, because of positivity, any
such point must lie inside a tetrahedron [Fig. \ref{fig.dynamics} (a)] of
vertices $(-1,-1,-1)^T$, $(-1,1,1)^T$, $(1,-1,1)^T$ and $(1,1,-1)^T$, which
represent the four Bell states \cite{Horodecki:1996gf}. Inside this tetrahedron
we distinguish an inner octahedron, which contains the separable states, from
the four remaining corners, which consist of the entangled states
\cite{Horodecki:1996gf} and are labeled by the Bell state they contain (e.g.,
$\ket{\Psi_-}$-corner). In this setting Wootters's concurrence \cite{Wootters} is
simply the distance from the faces of the octahedron: $C(\vb d) =1/2 \max
\lbrace 0 , \sum_i |d_i|-1 \rbrace$. Equidistant points, parallel to the
surfaces of the octahedron, form the ``isoconcurrence'' planes.

In the tetrahedron, the collective dephasing evolution is always constrained
onto a plane defined by $\mathrm{Tr}\beta(t)=\mathrm{Tr}\beta(0)$ \cite{Supp}.
In the $\ket{\Psi_-}$-corner, these planes coincide with isoconcurrence planes,
which implies that entanglement is preserved for all of these states, for
arbitrary directions of the magnetic field. This leads to a finite-measure set
of states with time-invariant concurrence, despite the fact that those states do
evolve in time, $\rho(t)\neq\rho(0)$ \cite{time-invariant}. For entangled states
outside the $\ket{\Psi_-}$-corner, we can use Eq.~\eqref{eq.timeevolution} to
predict the final concurrence as $C_{\vb d, \text{f}} ( \vb n ) = 1/2 \max
\lbrace 0 , \sum_{i=1}^3 (1-2n_i^2 ) d_i - 1 \rbrace$, where $\vb
n=(n_1,n_2,n_3)^T$ and $\vb d=(d_1,d_2,d_3)^T$ characterizes the initial state
\cite{Supp}. Thus, by solving for $\vb n$, we can always find a field direction
such that the entanglement is preserved at all times. This can be seen from the
long-time limit in Fig. \ref{fig.dynamics} (b), whereas the transient time
evolution depends on $p(\omega)$, as we will discuss later.

Collective interactions become particularly relevant in multipartite settings,
where decoherence and dissipation can be strongly enhanced
\cite{dicke,palma,lambropoulos00,Breuer}. To analyze the effect of collective
dephasing on multipartite entanglement, analytic expressions \textit{\`a la} Wootters
\cite{Wootters} are not available. Intricate hierarchies of multipartite
entanglement \cite{levi} can, however, be characterized efficiently by resorting
to separability criteria based on inequalities \cite{Guehne,Flo}. An $N$-partite
state $\rho$ is called $k$-separable if it can be written as a mixture of states
of the form $\rho = \rho_{A_1} \otimes \dots \otimes \rho_{A_k}$, where $A_1
\ldots A_k$ label a division of the $N$ parties into $k$ subgroups. For
instance, the matrix elements in an arbitrary basis of any $k$-separable
$N$-qubit density matrix $\rho$ satisfy $\sum_{0\leq i<j\leq N-1}\vert
\rho_{2^i+1,2^j+1}\vert \leq \sum_{0\leq i<j\leq N-1}\sqrt{\rho_{1,1}
\rho_{2^i+2^j+1,2^i+2^j+1}} + (N-k)/2 \sum_{i=0}^{N-1} \rho_{2^i+1,2^i+1}$
\cite{gao2,basis}. Defining $k_\text{eff}$ as the largest integer $k$ saturating
this inequality provides an upper bound to the state’s $k$-separability class,
as $k \leq  k_\text{eff}$. When $k_\text{eff} < 2$ the state certainly contains
genuine multipartite entanglement, i.e., it is not even 2-separable, while the
state can be fully separable ($N$-separability) only if $k_\text{eff} \geq N$.

\begin{figure}[tb]
\centering
\includegraphics[width=.48\textwidth]{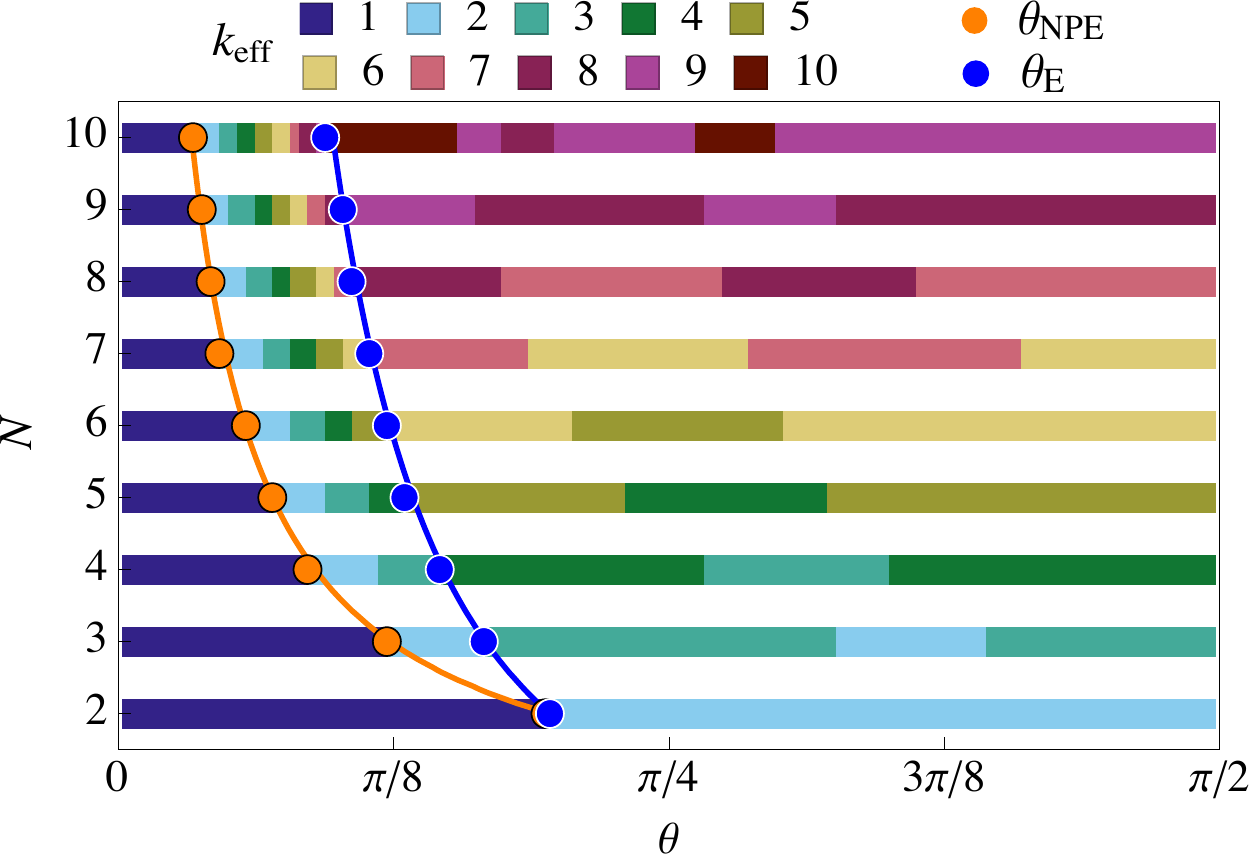} 
\caption{Influence of collective dephasing \eqref{dephasing_map}
on $N$-partite entangled $W$-states. Here we show the upper bound $k_\text{eff}$ to the asymptotic
state's separability vs the polar angles $0 \leq \theta \leq \pi/2$ (measured from
the $z$-axis to the $x,y$-plane) of the fluctuating magnetic field's direction.
In blue and orange we show, respectively, the dependence of the critical angles $\theta_\text{E}$ and $\theta_\text{NPE}$, Eqs. \eqref{eq:te} and \eqref{eq:tnpe}, on $N$.
The dots correspond to the numerical estimation of the smallest angle where $k_\text{eff}$
changes from 1 to 2 (defining $\theta_\text{NPE}$) or from $N-1$ to $N$
(defining $\theta_\text{E}$). The uncertainty on this angle, due to the finite
bin width of our sampling, is covered by the dot size. The lines represent the
expressions \eqref{eq:te} and \eqref{eq:tnpe}.}
\label{fig:multipartite}
\end{figure}

We consider the initial ($N$-partite entangled) $W$-state, $\ket W = (\ket{1 0
\dots 0} + \ket{0 1 \dots 0} + \dots + \ket{0 \dots 0 1})/\sqrt{N}$, where
$|1\rangle$ and $|0\rangle$ denote eigenstates of $\sigma_z$. Since the
collective dephasing map \eqref{dephasing_map} is invariant under the operation
$\vb n \rightarrow - \vb n$ and, additionally, this class of states exhibits
rotational symmetry around the $z$-axis, the polar angle $\theta\in [0,\pi/2]$
between $\vb n$ and the $z$-axis (which is defined by the local eigenbasis of
the initial state) fully determines the evolution of the state under
\eqref{dephasing_map}. Figure~\ref{fig:multipartite} displays the entanglement
properties of the resulting asymptotic state, characterized by
$k_{\mathrm{eff}}$ as a function of $\theta$. In general, there are relatively
small angle intervals that lead to a fully separable state, and typically
$k_{\mathrm{eff}}$ shows nonmonotonic dependence on $\theta$. 

Our numerical data (Fig.~\ref{fig:multipartite}) suggest that the asymptotic
state resulting from $|W\rangle$ is certainly entangled (i.e. $k_{\mathrm{eff}}<N$) as long as
$\theta<\theta_\text{E}$, where
\begin{equation}\label{eq:te}
\theta_\text{E} (N) = \arctan\left(1/\sqrt{N}\right).
\end{equation}
Conversely, when we choose a magnetic field that is close to the $z$-direction,
the initial $N$-partite entanglement of the $W$-state will be preserved during
the dephasing process, since $|W\rangle$ is part of an eigenspace of the Hamiltonian for
$\vb n = (0,0,1)^T$. Again, we find a critical angle
\begin{equation}\label{eq:tnpe}
\theta_\text{NPE} (N)= \arctan\left(1/\sqrt{N(N-1)}\right),
\end{equation}
such that for $\theta<\theta_\text{NPE}$, the asymptotic state will contain
genuine multipartite entanglement ($k_\text{eff} < 2$). Conditions \eqref{eq:te}
and \eqref{eq:tnpe} provide a finite range of orientations that ensure
preservation of entanglement properties in initial $W$-states. However, as the number of qubits gets
larger, higher accuracy is required to maintain $N$-partite entanglement
($\theta_\text{NPE}$) or at least some type of entanglement ($\theta_\text{E}$).
Moreover, the fast decay of $\theta_\text{NPE}$ with the number of qubits
confirms that genuine $N$-partite entanglement is much more fragile than
bipartite entanglement \cite{levi,Monz}, which is able to resist a larger range
of field directions. We remark here that in order to modify $\theta$ in a
trapped-ion experiment it is much more natural to apply unitary pulses to the
initial state to shift its relative orientation to the field, instead of
actually changing the orientation of the external field \cite{Ben}.

\begin{figure}[tb]
\centering
\includegraphics[width=.48\textwidth]{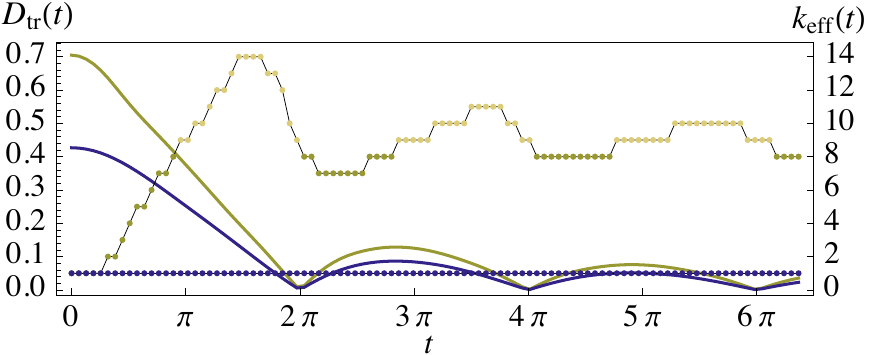} 
\caption{(Color online) Evolution of
the trace distance $D_\text{tr}$ between $\rho(t)$ and the asymptotic state
$\rho_s$ (solid lines), and of the state's separability bound $k_\text{eff}$
(connected dots), respectively, for eight-partite $|W\rangle$ (green) and $|\tilde W\rangle$ (blue) initial states (see the text for a definition), box-distributed noise fluctuations
$B_{0,1}(\omega)$, and a polar angle $\theta = \pi/8$.
The values where $k_\text{eff} > N$, marked in a lighter green shade, indicate that the state is compatible with full separability, and hence they do not provide additional information than the case $k_\text{eff} = N$. Note that the $|\tilde W\rangle$ state exhibits time-invariant genuine multipartite entanglement outside of a time-invariant subspace.}
\label{fig:multipartitedyn}
\end{figure}

Furthermore, we notice that states displaying time-invariant entanglement
properties can be found in the multipartite case, too. One example is given by a
specific family of $W$-states, whose single-excited states
carry the relative phases $\lbrace \E^{\im 2 \pi k / N} \rbrace_{k=1}^N$ in an
arbitrary order, e.g.,
\begin{align}
|\tilde W\rangle=& ( \E^{\im (2 \pi / N)}\ket{1 0
\dots 0} +\E^{\im (4 \pi / N)} \ket{0 1 \dots 0}  \nonumber \\
&  + \dots + \ket{0 \dots 0 1})/\sqrt{N}.
\end{align}
As shown in Fig. \ref{fig:multipartitedyn}, this state remains
$N$-partite entangled throughout the whole evolution, but the state itself
evolves into a stationary state, as is displayed by the trace distance
$D_\text{tr}(t)=\|\rho(t)-\rho_s\|/2$, where
$\|X\|=\mathrm{Tr}\sqrt{X^{\dagger}X}$ denotes the trace norm. The question
remains whether this state is part of a finite-measure set of states whose
multipartite entanglement properties are conserved, similarly to the
$\ket{\Psi_-}$-corner in the bipartite case---notice that $|\tilde W\rangle$ reduces to $|\Psi_-\rangle$ when $N=2$. Such states would constitute ideal
candidates for quantum computations by exhibiting invariance under collective
dephasing effects.

Let us finally characterize a family of time-invariant states, for arbitrarily
many qubits. Using Eq.~\eqref{eq.timeevolution}, it can be shown that any state
of the form $\rho_W=\sum_{s\in \Sigma_N}c_{s}V_{s}$, where $c_{s}$ are arbitrary
coefficients and $V_{s}$ are the permutation operators defined above,  satisfies
$\rho(t)=\rho(0)$ at all times \cite{Supp}. These states, known as multipartite
Werner states \cite{multiwerner}, are also characterized by their invariance
under arbitrary local unitary transformations $U^{\otimes N}$ \cite{weyl}. Since
such transformations describe collective changes of the local qubit coordinate
systems, it is quite intuitive that these states are time-invariant for
\emph{arbitrary directions} of the external field. This identifies a
$(N!-1)$-parameter family of states that always span a decoherence-free subspace
\cite{DFS,palma}. In the geometric picture of Fig. \ref{fig.dynamics} (a), these
states lie on the line passing through the origin of the tetrahedron and the
$\ket{\Psi_-}$-state.

We conclude with some remarks on the transient evolution towards the asymptotic
state. To determine how close the evolved state is to its asymptotic state, we
again employ the trace distance $D_\text{tr}(t)$ which has a clear
interpretation in terms of the distinguishability of the quantum states
\cite{Hayashi}. In our present context, the trace distance is employed as an
autocorrelation function, which reveals the monotonicity of the quantum
evolution.

While different types of noise fluctuations lead to the same asymptotic state,
as discussed earlier, the transient behavior can be qualitatively different, as
displayed in Fig.~\ref{fig.dynamics}. When the distribution $p(\omega)$ is
Lorentzian, $C_{\omega_0, \gamma} (\omega)= (\gamma/\pi) [(\omega - \omega_0)^2
+ \gamma^2]^{-1}$, or Gaussian, $N_{\omega_0, \sigma}(x) = \exp(-(\omega -
\omega_0)^2/2 \sigma^2)/\sqrt{2\pi \sigma^2}$ (as suggested in
Ref.~\cite{Monz}), the properties of the state, such as the concurrence, decay
exponentially towards their asymptotic value [Fig.~\ref{fig.dynamics} (b)]. When
we instead consider the box distribution over the interval $\left[0 , \omega_0
\right]$, i.e., $B_{0,\omega_0}(\omega) = \left[\Theta_H(\omega) -
\Theta_H(\omega-\omega_0) \right]/\omega_0$, where $\Theta_H(\omega)$ is the
Heaviside step function, we observe a nonmonotonic approach of the quantum
system to the asymptotic state (Fig.~\ref{fig.dynamics}). In fact, for this
distribution the characteristic function $\varphi(t)$ is proportional to
$\sin(\omega_0t)/t$, which asymptotically decreases on a significantly longer
time scale than the exponential decay characterizing the Lorentzian or Gaussian
distributions. This nonmonotonic behavior also implies that the
ensemble-averaged dynamics of noninteracting atoms in a fluctuating classical
field cannot be modeled by an effective Markovian environment for certain noise
distributions $p(\omega)$ \cite{lambropoulos00,BLP}. These frequency
fluctuations, therefore, take on the role of the environment's spectral density in a
standard open-system description of decoherence \cite{lambropoulos00,Breuer}.

To summarize, we have provided a model for the dephasing dynamics of a
collection of noninteracting atoms subject to a homogeneous external field of
fluctuating intensity. The effective environment-induced interactions are
described analytically by an exact solution in terms of a canonical Kraus map,
able to describe the time evolution of multipartite systems under arbitrary
intensity fluctuations. Our model applies to a variety of experiments in atomic
physics, and describes one of the dominant error sources for state-of-the-art
trapped-ion experiments. Complete theoretical control on transient as well as
asymptotic dynamics allows for the formulation of precise conditions for preserving
relevant quantities, such as entanglement, as well as for the identification of
families of states whose properties are completely insensitive to the direction of
the external field.

\textit{Acknowledgment}. M.G. thanks the German National Academic Foundation for their support.

\clearpage

\part{Supplementary Material}

\section{Trace preserving property of the collective dephasing map}

The operators $\Lambda_\pm = \frac{1}{2}(\idm_2 \pm \vb n \cdot \vb \sigma)$ form a complete set of orthogonal projectors on the Hilbert space $\mathcal{H} \simeq \mathbb{C}^2 $ of each qubit. These properties are inherited by the $\Theta_i$ operators, which are themselves orthogonal projectors: $\Theta_i = \Theta_i^\dagger$ and $\Theta_i \Theta_j = \Theta_i \delta_{i,j}$.

The trace preserving property of the map therefore reduces to $\sum_i \Theta_i^\dagger \Theta_i = \sum_i \Theta_i = \idm_{2^N}$. This is simply proven by putting $t=0$ in Eq. (2) from the main text, and following the equalities from left to right:
\begin{equation}
\sum_{i=0}^N \Theta_i = (\Lambda_+ + \Lambda_-)^{\otimes N} = \idm_{2^N}  .
\end{equation}

For the $A_i(t)$ operators we instead have
\begin{align}
\sum_{i=0}^N A_i^\dagger(t) A_i(t) & = \sum_i \sum_{j,k} \lambda^i(t) \lambda^i_j(t) [\lambda^i_k(t)]^* \Theta_i \Theta_j \nonumber \\
 & = \sum_i \sum_{j} \lambda^i(t) \lambda^i_j(t) [\lambda^i_j(t)]^* \Theta_j \nonumber \\
 & = \sum_j \Theta_j = \idm_{2^N}  ,
\end{align}
where we have used the spectral decomposition $M(t) = \sum_i \lambda^i(t) \vb \lambda^i(t){\vb \lambda^i}^\dagger(t)$, together with $M_{jj}(t) = 1, \forall t$.

\section{Conserved trace of the $\beta$ matrix}
We now prove that the trace of the $\beta(t)$ matrix, defined by $\beta_{ij} (t)= \tr \left[\rho(t) \cdot \sigma_i \otimes \sigma_j \right]$, is a time-invariant quantity.  From the definition we have
\begin{align}
\tr \beta(t) &= \sum_{i=1}^3 \beta_{ii}(t) = \sum_{i=1}^3  \tr \left(\rho(t) \cdot \sigma_i \otimes \sigma_i \right) \notag\\
&= \tr \left[ \rho(t) \sum_{i=1}^3 \sigma_i \otimes \sigma_i \right]  .
\end{align}
Notice now that the Bell state $\ket{\Psi^-} \bra{\Psi^-}$ reads \cite{Horodecki:1996gf}
\begin{align}
\ket{\Psi^-} \bra{\Psi^-} = \frac{1}{4} \left(\idm_4 - \sum_{i=1}^3 \sigma_i \otimes \sigma_i \right)  ,
\end{align}
which yields $\sum_{i} \sigma_i \otimes \sigma_i = \idm_4 - 4 \ket{\Psi^-} \bra{\Psi^-}$. Substituting back we then have
\begin{align}
\tr \beta(t) &= \tr \rho(t) - 4 \tr\left[ \rho(t) \ket{\Psi^-} \bra{\Psi^-} \right] \notag\\
&= 1 - 4\tr\left[ \rho(t) \ket{\Psi^-} \bra{\Psi^-} \right] .
\end{align}

To prove that $\tr \beta(t)$ is conserved under time evolution, we compute its derivative and check whether it vanishes:
\begin{align}
\frac{\dif \tr \beta(t)}{\dif t} = -4 \tr\left[ \dot \rho(t) \ket{\Psi^-} \bra{\Psi^-} \right]  .
\end{align}
The time derivative of $\rho(t)$ reads
\begin{align}
\dot \rho(t) = \mathcal{L}_t \left[\rho(t) \right] = \sum_{i,j=0}^N \dot M_{ij}(t) \Theta_i \rho(t) \Theta_j ,
\end{align}
which implies that
\begin{align}
\quad&\tr\left[ \dot \rho(t) \ket{\Psi^-} \bra{\Psi^-} \right] \notag\\& = \tr\left[  \sum_{i,j=0}^N \dot M_{ij}(t) \Theta_i \rho(t) \Theta_j\ket{\Psi^-} \bra{\Psi^-} \right]\notag \\
& =\tr\left[  \sum_{i,j=0}^N \dot M_{ij}(t) \Theta_j\ket{\Psi^-} \bra{\Psi^-}  \Theta_i \rho(t) \right] \notag\\
& = \tr\left[  \left( \mathcal{L}_t \left[ \ket{\Psi^-} \bra{\Psi^-} \right]\right)^\dagger\rho(t) \right]  .
\end{align}
However, the Bell state $\ket{\Psi^-}$ is an eigenstate of the Hamiltonian of the system for whichever choice of $\vb n$, which means that it is itself unaffected by collective dephasing. This implies that $\mathcal{L}_t \left[ \ket{\Psi^-} \bra{\Psi^-} \right] = 0, \forall t$ and therefore
\begin{align}
\frac{\dif}{\dif t} \tr \beta(t) \equiv 0  .
\end{align}

\section{Concurrence for Bell-diagonal states}
In the tetrahedron of Bell-diagonal states \cite{Horodecki:1996gf}, Wotters' concurrence \cite{Wootters} reads
\begin{align}
C(\vb d) = \frac{1}{2} \max\left\lbrace 0,-1+\sum_i \left| d_i \right| \right\rbrace  ,
\end{align}
where $\vb d = (d_1,d_2,d_3)^T \in \mathbb{R}^3$ is the point representing the quantum state (see main text).

For states in the $\ket{\Psi_-}$-corner, we have $d_i \leq 0, \forall i$. Hence, the concurrence in this corner can be rewritten as
\begin{align}
C(\vb d)& = \frac{1}{2} \max\left\lbrace 0,-1-\sum_i d_i \right\rbrace\notag\\& = \frac{1}{2} \max\left\lbrace 0,-1-k \right\rbrace  ,
\end{align}
where we have used the fact that $\tr \beta = \sum_i d_i = k$ is a constant. This explicitly proves that Bell-diagonal states in this corner have time-invariant concurrence.

In the other corners of the tetrahedron, only one of the coordinates is negative. If we suppose that $d_1 \leq 0$ (i.e. $\ket{\Phi_-}$-corner), we have
\begin{align}
-1+\sum_i \left| d_i \right| = -1 -d_1 + d_2 + d_3 = -1 +k -2d_1  .\notag
\end{align}
Let $\vb d$ represent the initial state, whereas $\vb d^\text{f}$ represents the final, asymptotic state of the system after collective dephasing. Furthermore, we denote the negative components of $\vb d$ and $\vb d^\text{f}$ with a subscript $j$, i.e., we have $d_j \leq 0$ and $d^\text{f}_j \leq 0$, respectively. Direct application of the collective dephasing map leads to
\begin{align}
d_j^\text{f} = \sum_i d_i n_i^2  .
\end{align}
The concurrence in the final state is then
\begin{align}
C \left(\vb d^\text{f}\right) &= \frac{1}{2} \max 	\left\lbrace 0 , -1 + \sum_i d_i - 2 \sum_i d_i n_i^2 \right\rbrace \notag\\
& = \frac{1}{2} \max 	\left\lbrace 0 , -1 + \sum_i (1 - 2 n_i^2) d_i \right\rbrace .
\end{align}
From these formulae one can immediately notice that $C(\vb d) = C(\vb d^\text{f}) \iff \vb n = \pm \vb e_j$, where $\lbrace \vb e_1,\vb e_2,\vb e_3 \rbrace$ is the standard basis of $\mathbb{R}^3$.

\section{Time-invariance of multipartite Werner states}
In this Section we prove that the multipartite Werner states \cite{multiwerner} are time-invariant under the action of the collective dephasing map. Let $s \in \Sigma_N$ be a permutation in the symmetric group, and $V_s$ its representation in the operator space of $N$ qubits:
\begin{equation}
V_{s}=\sum_{i_1\dots i_N \in \lbrace 0,1 \rbrace}|i_{s(1)}\dots i_{s(N)}\rangle\langle i_1\dots i_N|  .
\end{equation}
The multipartite Werner states are then defined as $\rho_W=\sum_{s\in \Sigma_N}c_{s}V_{s}$, where $c_s$ are arbitrary coefficients, leading to a valid quantum state $\rho_W$.

Analogously, if we define $k_i = i!(N-i)!$ and $\mathcal{Q}_i = \Lambda_+^{\otimes i} \otimes \Lambda_-^{\otimes N-i}$, we can rewrite the $\Theta_i$ operators as
\begin{equation}
\Theta_i = \frac{1}{k_i} \sum_{s \in\Sigma_N} V_s \mathcal{Q}_i V_s^{\dagger}.
\end{equation}

Direct application of the collective dephasing map yields
\begin{align}
\epsilon[\rho_W] = \sum_{i=0}^N \sum_{\pi,\sigma,\lambda \in \Sigma_N} \frac{c_\pi}{k_i^2}\left(V_\sigma 
 \mathcal{Q}_i V_\sigma^{\dagger} \right) V_\pi \left( V_\lambda  \mathcal{Q}_i V_\lambda^{\dagger} \right)  .
 \end{align}
Because $\Sigma_N$ is a closed group, the concatenation of two permutations describes another permutation, and therefore $\exists \alpha \in \Sigma_N : V_\sigma = V_\pi V_\alpha$. The expression above can then be rewritten as
\begin{align}
&\quad\epsilon[\rho_W] \notag\\
&= \sum_{i=0}^N \sum_{\pi,\alpha,\lambda \in \Sigma_N} \frac{c_\pi}{k_i^2} V_\pi \left(V_\alpha 
 \mathcal{Q}_i V_\alpha^{\dagger} \right)\left( V_\lambda  \mathcal{Q}_i V_\lambda^{\dagger} \right) \notag\\
 & = \sum_{\pi \in \Sigma_N} c_\pi V_\pi \sum_{i=0}^N \sum_{\alpha \in \Sigma_N} \frac{\left(V_\alpha 
 \mathcal{Q}_i V_\alpha^{\dagger} \right)}{k_i} \sum_{\lambda \in \Sigma_N} \frac{\left( V_\lambda  \mathcal{Q}_i V_\lambda^{\dagger} \right)}{k_i}\notag \\
 & = \sum_{\pi \in \Sigma_N} c_\pi V_\pi \sum_{i=0}^N \Theta_i \Theta_i  = \rho_W  ,
 \end{align}
where we have used the idempotency of the $\Theta_i$ operators and the closure relation $\sum_i \Theta_i = \idm_{2^N}$.


\begin{thebibliography}{xx}
\bibitem{abscont} These are just the ``usual'' continuous probability
distributions $f$ that can be defined via (Lebesgue-integrable) 
cumulative distribution functions $F$ as $F(x) = \int_{-\infty}^x f(a) \dif a$.
The Riemann-Lebesgue lemma \cite{rudin2}
then ensures that the characteristic function vanishes asymptotically (see
Ref.~\cite{lukacs}).
\bibitem{basis} Here the density matrix is represented in the standard product
basis of the eigenvectors of $\sigma_z$ \cite{Guehne}.
\end{thebibliography}
\end{document}